\documentclass[aps,prb,showpacs,reprint, superscriptaddress]{revtex4-1}

\usepackage{graphicx}
\usepackage{color}
\usepackage{amsmath}
\usepackage{bbm}
\usepackage{amssymb}

\usepackage{dsfont}

\usepackage[utf8]{inputenc}	
\usepackage[T1]{fontenc}

\input epsf

\begin{document}

\title{Superconducting properties of the hole-doped three-band \emph{d-p} model studied with minimal-size real-space \emph{d}-wave pairing operators}

\author{A. Biborski}
\email{andrzej.biborski@agh.edu.pl}
\affiliation{Academic Centre for Materials and Nanotechnology, AGH University of Science and Technology, Al. Mickiewicza 30, 30-059 Krakow,
Poland}
\author{M. Zegrodnik}
\email{michal.zegrodnik@agh.edu.pl}
\affiliation{Academic Centre for Materials and Nanotechnology, AGH University of Science and Technology, Al. Mickiewicza 30, 30-059 Krakow,
Poland}
\author{J. Spa\l ek}
\email{jozef.spalek@uj.edu.pl}
\affiliation{Institute of Theoretical Physics, 
Jagiellonian University, ul. \L ojasiewicza 11,
30-348 Krakow, Poland}

\date{24.04.2020}

\begin{abstract}
The three-band \emph{d-p} model is investigated by means of Variational Monte-Carlo (VMC) method with the BCS-like wave-function supplemented by the Gutzwiller and Jastrow correlators. The VMC optimization leads to $d$-$wave$ superconducting state with a  characteristic dome-like shape of the order parameter for hole doping $\delta \lesssim 0.4$, in a good agreement with the experimental observations. Also, the off-diagonal pair-pair correlation functions, calculated within VMC, vindicates
the results obtained very recently within the diagrammatic expansion of the Gutzwiller wave function method (DE-GWF) [cf. Phys. Rev. B \textbf{99}, 104511 (2019)]. Subsequently, the nature of the $d$-$wave$ pairing is investigated by means of recently proposed \emph{minimal-size real-space d-wave pairing operators} [Phys. Rev. B \textbf{100}, 214502 (2019)]. An emergence of the long-range superconducting ordering for both $d$ and $p$ orbitals is reported by analysing the corresponding off-diagonal pair-pair correlation functions. The dominant character of \emph{d-wave} pairing on $d$ orbitals is confirmed. Additionally, the trial wave-function is used to investigate the magnetic properties of the system. The analysis of spin-spin correlation functions is carried out and shows antiferromagnetic $\mathbf{q}=(\pi,\pi)$, short-range order, as expected. For the sake of completeness, the charge gap has been estimated, which for the parent compound takes the value $\Delta_{CG}\approx1.78\pm0.51\text{ eV}$, and agrees with  values reported experimentally for the cuprates. 
\end{abstract}

\maketitle
\section{Introduction}
The unconventional superconductivity discovered in copper based compounds by Bednorz and Müller in 1986 is still under intensive debate~\cite{KEREN}. This class of systems is difficult to handle realistically by means of the most popular quantum chemistry method, i.e., density functional theory (DFT), due to the fact that the electron-electron interactions play a crucial role in  the resulting physical properties. As electronic correlations cannot be described consistently within any known mean-field formalism (e.g. \emph{double counting} problem in DFT methods), simplified models, capturing the essentials of electronic structure are required. The application of the cannonical single-band models used for recaption of the correlated systems (Hubbard and $t$-$J$ models\cite{Hubbard,tJmodel_rev_2008}) allowed for the reproduction of both the Mott insulating phase at half-filling and the superconducting state for the electron- and hole- doped cases. In such approaches the initially multi-band problem ($d$-$p$ model) is mapped onto a single-band picture in which the Zhang-Rice singlets \cite{ZhangRice} play the role of \emph{quasiparticles}. It is believed that many of the unusual properties of the cuprates arise from the electronic degrees of freedom of the copper-oxygen planes, which are common for the whole cuprate family.  Although the mentioned models allow to reproduce the selected fundamental features of the cuprates, other subtle phenomena such as charge(spin)-density-waves or nematicity appearance, may directly emerge from  the interplay between \emph{d} and \emph{p} orbitals ~\cite{Comin}. However, the question of the minimal model which captures the cuprate physics to a satisfactory extent still remains an open issue and ongoing research of both single- and multiband- approaches is at place. With this respect, it is worth noting that the description of the mentioned ordered phases within the single band picture has recently lead to some interesting results ~\cite{Jiang1424, Zheng1155,IdoKota,Zegrodnik_5}. Nevertheless, microscopic insight into the pairing between $d$-$d$, $p$-$d$ and $p$-$p$ channels can lead to better understanding of the superconducting state\cite{Moreo}, as suggested by some of the experimental observations\cite{Rybicki2016,STARFISH}. Therefore, it is natural to consider  more realistic model in which, the unit cell consists of one \emph{d} orbital and two \emph{p} orbitals. Regardless of the number of bands considered, exact ground state for Hubbard-type Hamiltonians (excluding selected 1-d cases) is not known. Therefore, approximate methods are to be used in its diagonalization procedure. Whereas Exact Diagonalization (ED) techniques provide accurate numerical solution, they are limited to small systems, which essentially, cannot give answers related with the presence of the long-range electronic correlations. The state-of-art Density Matrix Renormalization Group (DMRG) method, though computationally demanding, has been profitably exploited for studying both charge order as well as pairing in strongly correlated model systems~\cite{Jiang1424, Zheng1155}. The Determinant Monte-Carlo (DMC) calculations, despite the infamous \emph{sign problem} are promising for the description of the cuprates\cite{QMC_3band_2016}. However, the paired state has not been explicitly included in such analysis. Another choice is the application of \emph{variational} methods which may be considered  as  well balanced in view of  their complexity and reliability of the obtained results. Therefore, a properly constructed trial wave function allows to gain insight into the nature of the ground state of the particular correlated electronic system\cite{Becca}.
 
 Encouraged by the results for the superconducting and nematic states obtained by means of the Diagrammatic Expansion of the Gutwziller Wave Function (DE-GWF method) approach for the three-band $d$-$p$ model\cite{ZEG2,ZEG1}, we have decided to characterize the superconducting properties, particularly in view of the spatial dependence of the correlation functions obtained by means of the Variational Monte-Carlo (VMC) calculation scheme. Numerous studies regarding this topic have been carried out up to now~\cite{VMC1,Dopf,Scalettar,HuangLin,QMC_3band_2016,SchwarzLauretta,ZEG1,Yanagisawa_2019,Batista,Asahata,Dopf2} related to  both \emph{normal} and \emph{superconducting} states. Here, we extend the  analysis of the SC state with an explicit calculations of the \emph{minimal-size real-space d-wave pairing operators} proposed very recently by Moreo and Dagotto \cite{Moreo}. To the best of our knowledge, their equal-time correlation functions have not been analyzed so far. We also supplement our analysis of the variational \emph{ansatz} for the paired state within the $d$-$p$ model with the intersite Jastrow-type correlators. 
 
  In the next Section we describe the model and sketch the method. Subsequently, in Sec. III we present the characteristics of the \emph{d-wave} superconducting phase for the hole-doped case by means of the \emph{standard} investigation, i.e., by analyzing the correlation functions for the \emph{d-wave} pairing between holes residing on the nearest-neighbor $d$ orbitals. Subsequently, we continue our analysis  of the correlation functions defined for the \emph{minimal-size real-space d-wave pairing operators}, consisting of a proper combination of the $d$-$p$ and $p$-$p$ pairing amplitudes. In Sec. IV, we also provide the spin-spin correlation functions and show the development of the short-range antiferromagnetic order, as well as  determine the value of the charge transfer gap. We conclude our results in the last Section.

\section{Three-band $d$-$p$ model and method}
\begin{figure}
 \centering
 \includegraphics[width=0.30\textwidth]{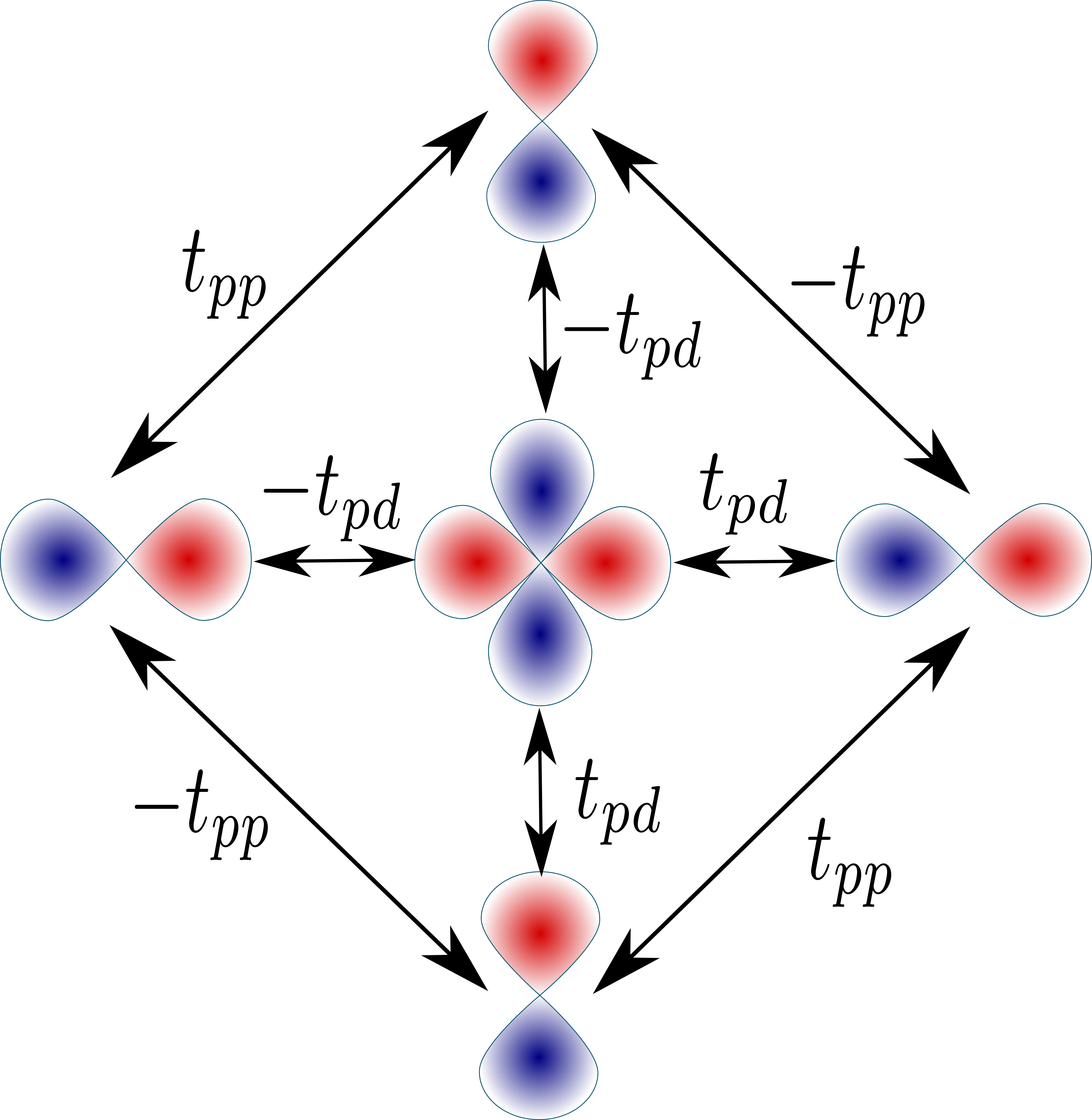}
 \caption{The hopping parameters included in three-band $d$-$p$ model described by  Hamiltonian defined in Eq.(\ref{eq:Hamiltonian_start}). Central orbital is $d_{x^2-y^2}$, states for to the copper atom whereas, remaining one are the oxygen $p_x$/$p_y$ orbitals.}
 \label{fig:Cu_O_hoppings}
\end{figure}

We consider the three-band $d$-$p$ model described by the Hamiltonian
\begin{equation}
\begin{split}
 \hat{\mathcal{H}}&=\sum_{\langle il,jl'\rangle}t^{ll'}_{ij}\hat{c}^{\dagger}_{il\sigma}\hat{c}_{jl'\sigma}+\sum_{il}\epsilon_{l}\hat{n}_{il}+\sum_{il}U_{l}\hat{n}_{il\uparrow}\hat{n}_{il\downarrow},\\ 
 \end{split}
 \label{eq:Hamiltonian_start}
\end{equation}
where $\hat{c}^{\dagger}_{il\sigma}$ ($\hat{c}_{il\sigma}$)  are creation (anihilation) fermionic operators acting on orbital $l \in \{d_{x^2-y^2},p_x,p_y\}$  related to $i$-th unit cell. As in our previous works, \cite{ZEG1,ZEG2} hoppings are limited to the nearest-neighboring orbitals (cf. Fig. \ref{fig:Cu_O_hoppings}). The values of hopping amplitudes $t_{ij}^{ll'}$ as well as atomic energy levels are set to, $t_{pp}=0.49$ eV, $t_{pd}=1.13$ eV, $\epsilon_p=-3.57$ eV and $\epsilon_d=0$, which are typical values for the cuprates~\cite{tJmodel_rev_2008}. The repulsive intra-orbital Hubbard interactions are $U_{p_x}=U_{p_y}=4.1$ eV   and $U_{d_{x^2-y^2}}=10.3$ eV  for oxygen and copper orbitals, respectively. This set of microscopic parameters was utilized recently in our DE-GWF solution ~\cite{ZEG1}, and, their values are similar to those applied by Kung et al\cite{QMC_3band_2016}.\\

As mentioned, we employ VMC approach in this study. This method exhibits both advantages as well as drawbacks which are common for the whole family of variational methods. Particularly, the proper choice of wave-function \emph{ansatz} is crucial to obtain reasonable output. Thus, one cannot expect to obtain properties of the system which are not encoded in the variational wave-function since the solution is narrowed to the sub-space in the Hilbert space. Therefore, the specific form of the wave-function  should be carefully chosen. This issue  does not appear in  the determinant quantum Monte Carlo (DQMC) which in principle (disregarding controllable approximations ) provides an \emph{exact} solution for the non-zero temperature. However, DQMC suffers for the infamous \emph{sign problem} and is more complex from the algebraic perspective. It must be stressed out that the variational wave-function in the VMC approach can be formulated in the manner which ensures reasonable computational costs and flexibility which is necessary to describe correlated system in many cases. Also, one of the main advantages of VMC in the  context of strongly correlated systems is its \emph{universality}, e.g., long-range interactions, as wel as three- and four-center two body terms can be encompassed almost effortlessly,  when the dose of generality is applied during the process of code development.\\
Our many-body trial wave-function is taken in the following manner\cite{Becca,MISAWA2019447}
\begin{equation}
|\Psi_T\rangle\equiv\hat{P}_G\hat{P}_J\hat{\mathcal{L}}_{S^{z}_{tot}}\hat{\mathcal{L}}_{N_{e}}|\Psi_0\rangle,
    \label{eq:trial_wf}
\end{equation}
where $\hat{P}_G$ is the Gutzwiller-type correlator given in the form
\begin{equation}
\hat{P}_G \equiv \text{exp}{\left[ -\sum_{l} g_{l}\sum_{i}{ }\hat{n}_{il\uparrow}\hat{n}_{il\downarrow} \right ]},
    \label{eq:p_gutzwiller}
\end{equation}
with $g_{p_x}=g_{p_y}$ due to the equivalency of the oxygen orbitals. The inter-orbital correlations are captured by the symmetric Jastrow density-density correlator
\begin{equation}
\hat{P}_J \equiv \text{exp}{\left[-\sum_{il,jl'}\lambda_{il,jl'}\hat{n}_{il}\hat{n}_{jl'}\right ]}.
    \label{eq:p_jastrow}
\end{equation}
Both $\{g_l\}$  and $\{\lambda_{il,jl'}\}$ are the subsets of variational parameters. When performing calculations for the $z$-component of the total spin, we set $S^z_{tot}=0$ and constant number of electrons  $N_{e}$, as well as the projectors $\hat{\mathcal{L}}_{S^{z}_{tot}}$ and $\hat{\mathcal{L}}_{N_{e}}$ are applied during sampling procedure. The non-interacting part $|\Psi_0\rangle$ is constructed from eigen-states of the BCS variational Hamiltonian $\hat{\mathcal{H}}_{eff}$ defined as
\begin{equation}
\begin{split}
 \hat{H}_{eff}=&\sum_{\langle il,jl'\rangle}\tilde{t}^{ll'}_{ij}\hat{c}^{\dagger}_{il\sigma}\hat{c}_{jl'\sigma}+\sum_{il}(\tilde{\epsilon_{l}}-\tilde{\mu})\hat{n}_{il}+\\ +&\sum_{il,jl'}\tilde{\Delta}^{ll'}_{ij}\hat{c}^{\dagger}_{il\uparrow}\hat{c}^{\dagger}_{jl'\downarrow}+h.c.
 \end{split}
 \label{eq:Hamiltonian_var}
\end{equation}
Note, that parameters with tilde,  are \emph{different} than those in non-interacting part of  $d$-$p$ Hamiltonian (Eq.~\ref{eq:Hamiltonian_start}) as they are considered as variational parameters to be optimized. More precisely, the above effective Hamiltonian defines the uncorrelated wave function $|\Psi_{0}\rangle$.
 The choice of both the hopping terms, and the pairing amplitudes is thus identical as in our DE-GWF study \cite{ZEG1}, i.e., it allows for the emergence of \emph{d-wave} pairing. It is worth to mention that during preliminary studies we have analyzed both \emph{d-} and \emph{s-wave}  pairing scenarios  within the DE-GWF approach, and, the \emph{d-wave} turned out to be the stable one. Therefore, we have $\Delta_{i+a_x,j}^{dd}=-\Delta_{i,j+a_y}^{dd}$, where $a_x/a_y=a$ refers to the nearest neighbour (nn) $d$ orbital in $\mathbf{x}$ and $\mathbf{y}$ directions, respectively. The diagonalization of Hamiltonian (\ref{eq:Hamiltonian_var}), which in turn allows to compose the many-electron part $|\Psi_0\rangle$; is divided into two stages. First, the following transformation of creation(anihilation) operators is applied. The spin-down-sector is converted to the hole picture, i.e., $\hat{c}_{il,\downarrow}^{\dagger}\rightarrow \hat{f}_{il,\downarrow}$ and $\hat{c}_{il,\downarrow}\rightarrow \hat{f}_{il,\downarrow}^{\dagger}$. Whereas the spin-up-sector operators are subject to the identity transformation, i.e.,  $\hat{c}_{il,\uparrow}^{\dagger}(\hat{c}_{il,\uparrow})\rightarrow \hat{f}_{il,\uparrow}^{\dagger}(\hat{f}_{il,\uparrow})$. The spin-down-sector transformation leads to the form of variational Hamiltonian, which can be directly diagonalized\cite{Becca}, by finding (numerically) the  unitary transformation for the  $N$-orbital system. \\ In effect, $|\Psi_{0}\rangle$ is defined in the standard manner, namely
\begin{equation}
    |\Psi_0\rangle_{\{\tilde{t}_{ij}^{ll'},\tilde{\epsilon_l},\tilde{\mu},\tilde{\Delta}_{il}^{ll'}\}}=\prod_{p=1}^{p=\tilde{n}}\hat\gamma_{p}^{\dagger}|\tilde{0}\rangle,
\label{eq:psi_0_var}
\end{equation}
where $|\tilde{0}\rangle$ is the  vacuum state for operators $\hat\gamma_{p}^{\dagger}$( $\hat\gamma_p$) representing quasi-particles, for which the variational Hamiltonian $\hat{\mathcal{H}}_{eff}$ can be written in the diagonal form. Note, that index $p=1,2,...\tilde{n}$ runs over first $\tilde{n}$ single-particle eigenstates of the variational Hamiltonian with $\tilde{n}=N+\sum_{il}(\hat{n}_{il\uparrow}-\hat{n}_{il\downarrow})$, resulting directly from the particle-hole transformation for the down-spins. The sampling procedure is executed in the standard manner. Configurations representing the distribution of $\tilde{n}$ particles among $N$ orbitals, $\{|x\rangle\}$, are sampled by means of Metropolis-Hastings \cite{Becca} algorithm according to the probability density $\rho(x)\propto |\langle x|\Psi_T\rangle|^2$.
 Physical quantities related to operators $\{\hat{O}\}$ are estimated as an average of their so called \emph{local values}\cite{Becca} $O_{loc}(x)$
\begin{equation}
    \langle \hat{O} \rangle\approx\frac{1}{M}\sum_{m=1}^M \frac{\langle x_m|\hat{O}|\Psi_T\rangle}{\langle x_m|\Psi_{T}\rangle}\equiv\sum_{m=1}^{M}O_{loc}(x_m),
\end{equation}
with $|x_m\rangle$ generated with respect to the probability density $\rho(x)$. In particular, the expectation value of the system energy, i.e., $\langle \hat{\mathcal{H}}\rangle$ can be computed for a given set of variational parameters. 
At least two, commonly exploited strategies for the wave-function optimization exist: \emph{variance optimization } and energy optimization. Sorella et al. elaborated the efficient procedure - \emph{Stochastic Reconfiguration} (SR) method \cite{Becca}, which benefits in  simultaneous optimization steps for the whole set of variational parameters. We have implemented the SR-based approach in our self-developed code (recently used also in a different context\cite{Biborski}) as it is regarded the state of art method in the field of interest \cite{MISAWA2019447}.

\section{Results}

In our computations the system is represented by the square cluster containing  $L\times L= 64$ unit cells, each consisting of one $d$-orbital and two $p$-orbitals ($p_x$ and $p_y$). This results in $64$ copper and $128$ oxygen atoms represented by appropriate orbitals. For the sake of clarity we define doping parameter $\delta$
\begin{equation}
    \delta\equiv 5-\frac{N_{e}}{L^2},
\end{equation}
i.e., the parent compound refers to $\delta=0$ with $5$ electrons per $\text{CuO}_2$ complex, and $\delta > 0$ corresponds to the hole-doped complex with $N_{e}<5$. We assume  $S^{z}_{tot}=0$; therefore minimal doping resolution is $\Delta\delta$=$2/64 = 0.03125$. The trial wave function is minimized with respect to the set of variational parameters by means of the SR method, and probed averages $\langle \hat{O} \rangle$ are sampled within $M\propto 10^7$ MC steps. Also, since VMC operates in the real space representation and the considered cluster is \emph{finite}, we apply the periodic boundary conditions. \\

\subsection{Superconducting correlations}

Within the VMC approach the superconducting properties of the system are typically determined by analysing the appropriate anomalous \emph{correlation functions} (CFs) (equal-time two-body Green functions). However, the choice of CFs is not unique and one may find particular form more suitable than other in the given methodological context \cite{Moreo}. First, we analyze the pairing between two $d$-$d$ holes by means of \emph{standard} equal-time CFs\cite{YAMAJI1998225,Dopf,Scalettar,Moreo} commonly used in the analysis of superconducting state in  real space. This part is regarded as validation of the applied method in view of our earlier DE-GWF solution\cite{ZEG1}. Next, we have applied the recently proposed \cite{Moreo} \emph{minimal-size real-space d-wave pairing operators} (MSPO) by means of their spatial dependency of CFs, to determine the \emph{d-wave} pairing properties within the three-band $d$-$p$ model. Specifically, these pairing operators refer to the possibility of intra-$p$ Cooper pairs formation \cite{Moreo,Batista,Littlewood}.
\subsubsection{Standard correlation functions}
\begin{figure}
 \centering
 \includegraphics[width=0.30\textwidth]{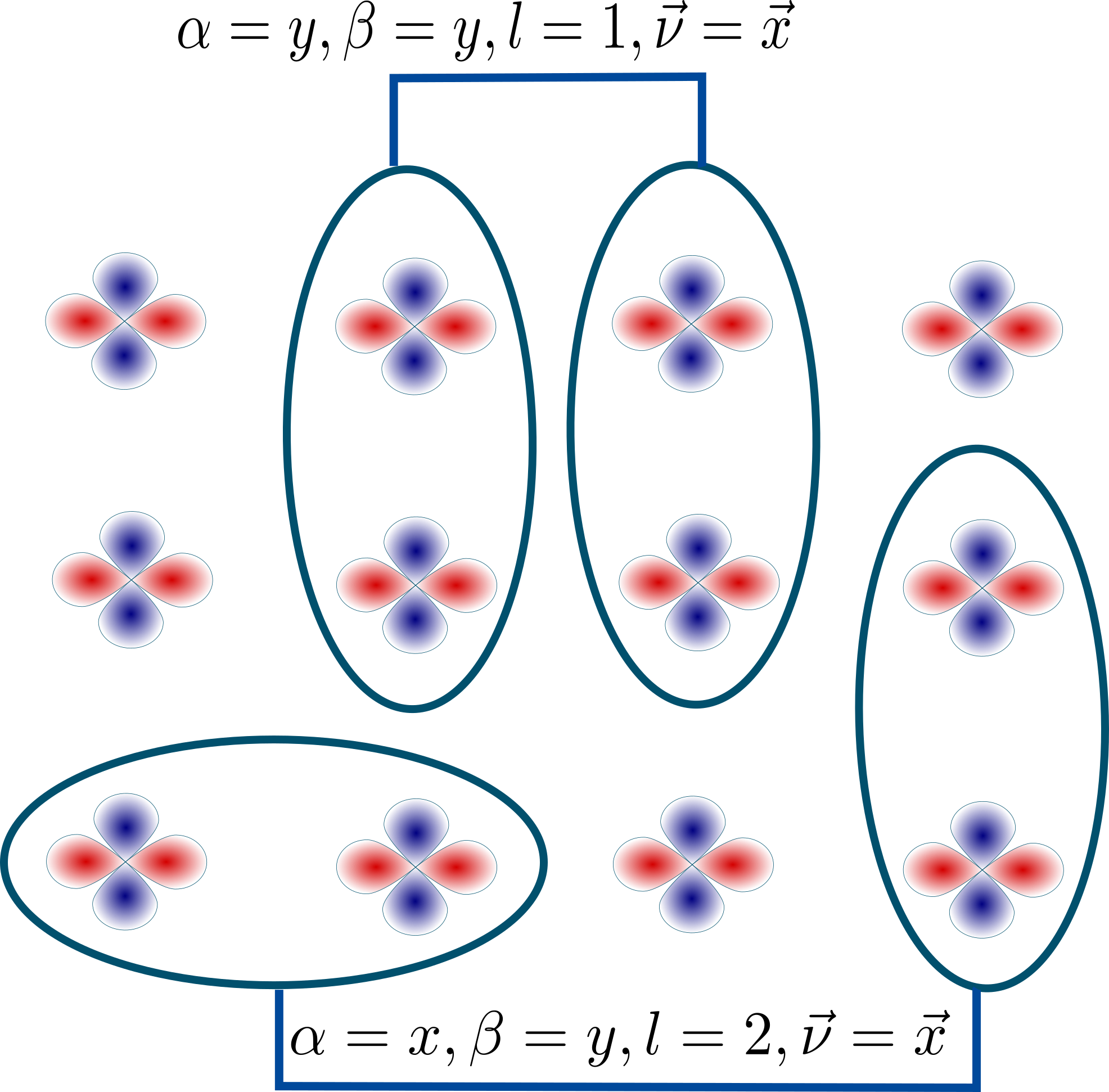}
 \caption{Schematic representation of  exemplary $d$-$d$ \emph{pair-pair} terms present in the correlation functions   defined in (\ref{eq:sc_correl}). }
 \label{fig:corr_fun}
\end{figure}

\begin{figure}
    \centering
    \includegraphics[width=0.5\textwidth]{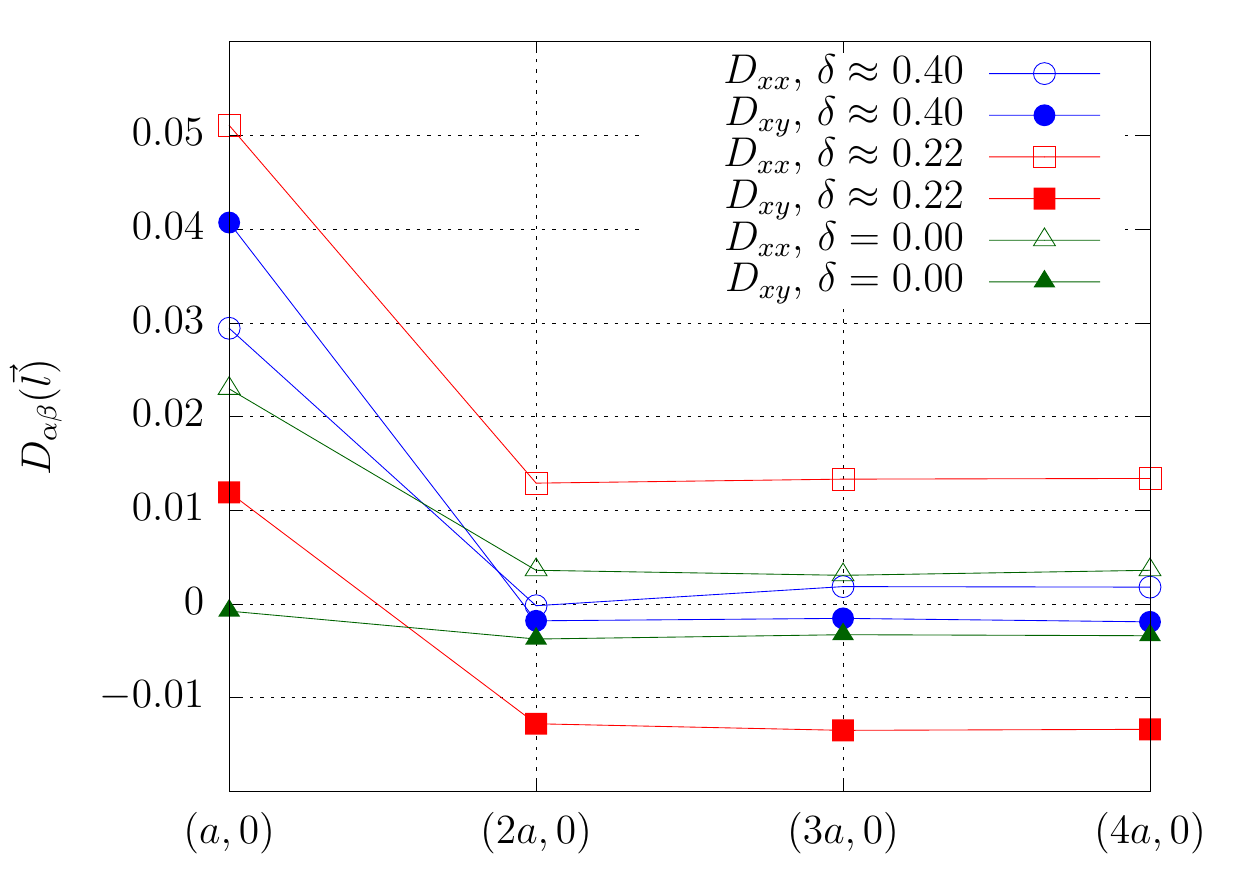}
    \caption{Correlation functions defined in (\ref{eq:sc_correl}), for the three representative hole-dopings and with $\mathbf{R}$ parallel to $\mathbf{x}$ direction.}
    \label{fig:corr_sc_spat}
\end{figure}

\begin{figure}
 \centering
 \includegraphics[width=0.5\textwidth]{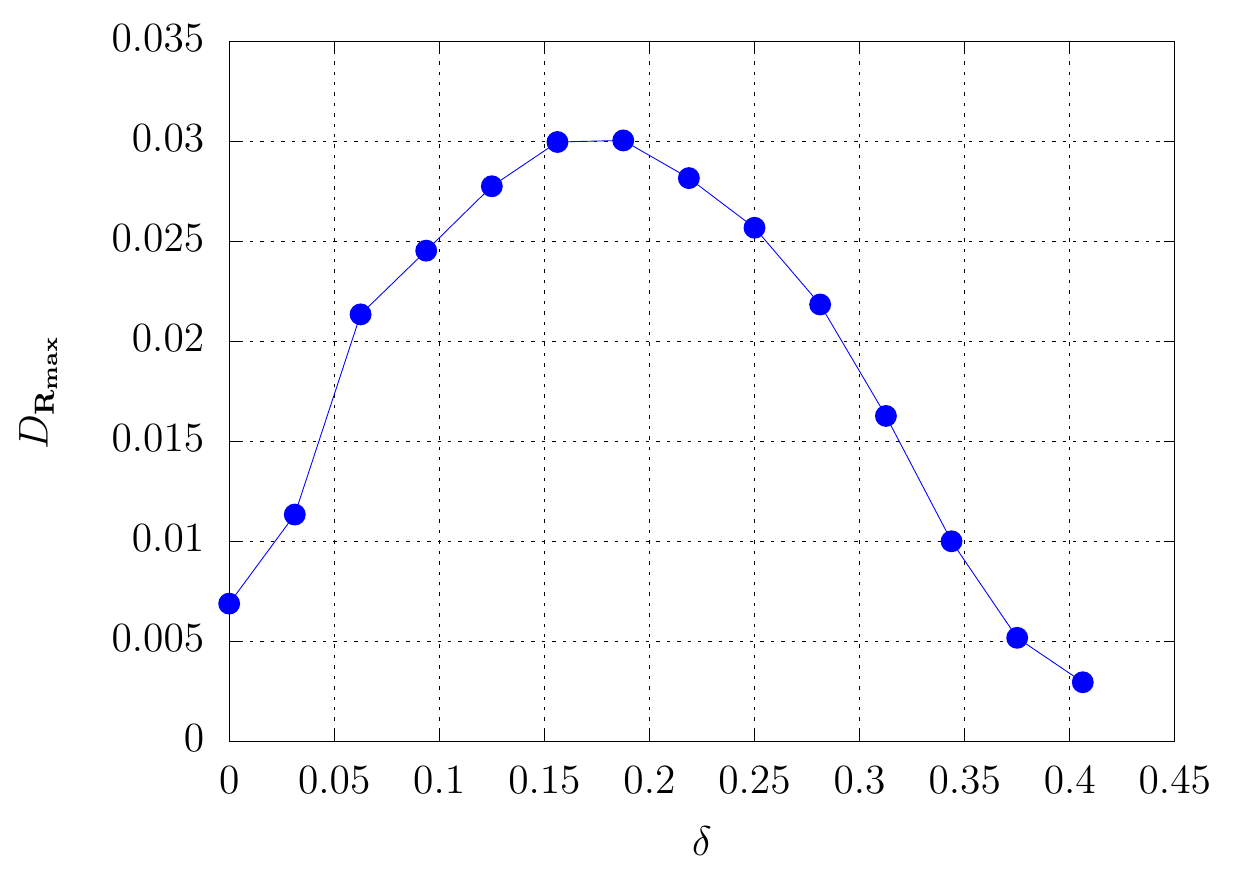}
 \caption{Superconducting order parameter $D_{\mathbf{R_{max}}}$, the line is guide for the eye.}
 \label{fig:sc_dome}
\end{figure}

To inspect fundamental  superconducting properties, as well as to  compare the results obtained by means of VMC with those of DE-GWF solutions, we analyzed first the spatial dependence of \emph{standard} off-diagonal pair-pair CFs for $d$-$d$ pairs, which is defined as
\begin{equation}
   D_{\alpha\beta}^{dd}(\mathbf{R})\equiv \frac{1}{L^2}\sum_{\mathbf{r}}\langle \hat{\Delta}_{\alpha}^{\dagger}(\mathbf{r}+\mathbf{R})\hat{\Delta}_{\beta}(\mathbf{r})\rangle,
  \label{eq:sc_correl}
\end{equation}
with $\alpha,\beta \in \{x,y\}$, and,
\begin{equation}
    \hat{\Delta}_{\alpha}^{\dagger}(\mathbf{r})\equiv\frac{1}{\sqrt{2}}\left (\hat{c}_{i(\mathbf{r})d\uparrow}^{\dagger}\hat{c}_{j(\mathbf{r}+\mathbf{a_{\alpha}})d\downarrow}^{\dagger}-\hat{c}_{i(\mathbf{r})d\downarrow}^{\dagger}\hat{c}_{j(\mathbf{r}+\mathbf{a_{\alpha}})d\uparrow}^{\dagger}\right).
\end{equation}
Function $i(\mathbf{r})$ maps the position of the center of the given orbital $d$ onto the index $i$. Vectors $\mathbf{a}_\alpha$ are given as
\begin{equation}
    \mathbf{a}_x=\begin{pmatrix}a \\ 0\end{pmatrix}, \mathbf{a}_y=\begin{pmatrix}0 \\ a\end{pmatrix},
\end{equation}
where $a$ is the lattice parameter.
The functions defined in Eq.(\ref{eq:sc_correl}) describe the spatial distribution of anomalous \emph{pair-pair} correlations, where each pair consists of two $d$-orbitals separated by the lattice constant $a$ in $\mathbf{x}$ or $\mathbf{y}$ directions  (c.f. Fig.~\ref{fig:corr_fun}). Note that as we analyze pure \emph{d-wave} pairing, the relation $D^{dd}_{\alpha\beta}(\mathbf{R})=D^{dd}_{\beta\alpha}(\mathbf{R})$ holds. It also should be mentioned that maximal distance refers to $\mathbf{R}_{max}=\left(\frac{L}{2},\frac{L}{2}\right)$ as we apply the periodic boundary conditions to the system.
In Fig.\ref{fig:corr_sc_spat} we present the spatial dependence of $D_{\alpha\beta}^{dd}$ for the selected  direction $\mathbf{R}\parallel \mathbf{x}$. It comes out that $D_{\alpha\alpha}^{dd}\approx-D_{\alpha\beta}^{dd}$ within the limit of attainable distance. Moreover, the values for $|\mathbf{R}|\geq 2a$, approach saturation, though for the high doping regime, correlations do not decay to zero (within the statistical error $\propto10^{-3}$). However, in the accessible maximal distance, we obtain very good agreement when compared to our recent analysis ~\cite{ZEG1}. In Fig.~\ref{fig:sc_dome} we present superconducting order parameter for the \emph{d-wave}-pairing defined as 
\begin{equation}
D_{\mathbf{R_{max}}}\equiv \sum_{\alpha\beta}(-1)^{1-\delta_{\alpha\beta}}D^{dd}_{\alpha\beta}(\mathbf{R=R_{max}}).
    \label{eq:sc_order}
\end{equation}
 We obtain a qualitative agreement when compared to the DE-GWF solution~\cite{ZEG1}, namely, the maximal amplitude of the order parameter appears at $\delta\approx0.15-0.2$. As already mentioned, the non-zero amplitude is present for each considered doping. This fact is due to a slow convergence of variational parameters in the high hole-doping regime, and/or, related to the limited cluster dimension.
Contrary to the DE-GWF solution we find it less problematic to optimize   wave function for  $\delta$ in the vicinity of the parent compound. In spite of $D(\delta=0)>0$, an abrupt decrease of the order parameter for $\delta\lessapprox0.1$ occurs and the obtained values of the order parameter form the \emph{dome}-like shape as a function of $\delta$, characteristic of the cuprates family.
We compare both methods quantitatively by computing the expectation values of nn. $(i,j)$ $d$-orbital pairs, namely
\begin{equation}
\Delta_{dd} \equiv \langle \hat{c}_{id\uparrow}^{\dagger}\hat{c}_{jd\downarrow}^{\dagger}\rangle,
\end{equation}
which is the measure of the superconducting order in the infinite system size, i.e., when
\begin{equation}
|\langle \hat{c}_{i(\mathbf{r})d\uparrow}^{\dagger}\hat{c}_{j(\mathbf{r}+\mathbf{a})d\downarrow}^{\dagger}\rangle|^2\approx \lim_{|\mathbf{R}|\rightarrow \infty}\langle\hat{\Delta}^{\dagger}({\mathbf{r}+\mathbf{R})}\hat{\Delta}({\mathbf{r})}\rangle.
\end{equation}

\begin{figure}
 \centering
 \includegraphics[width=0.5\textwidth]{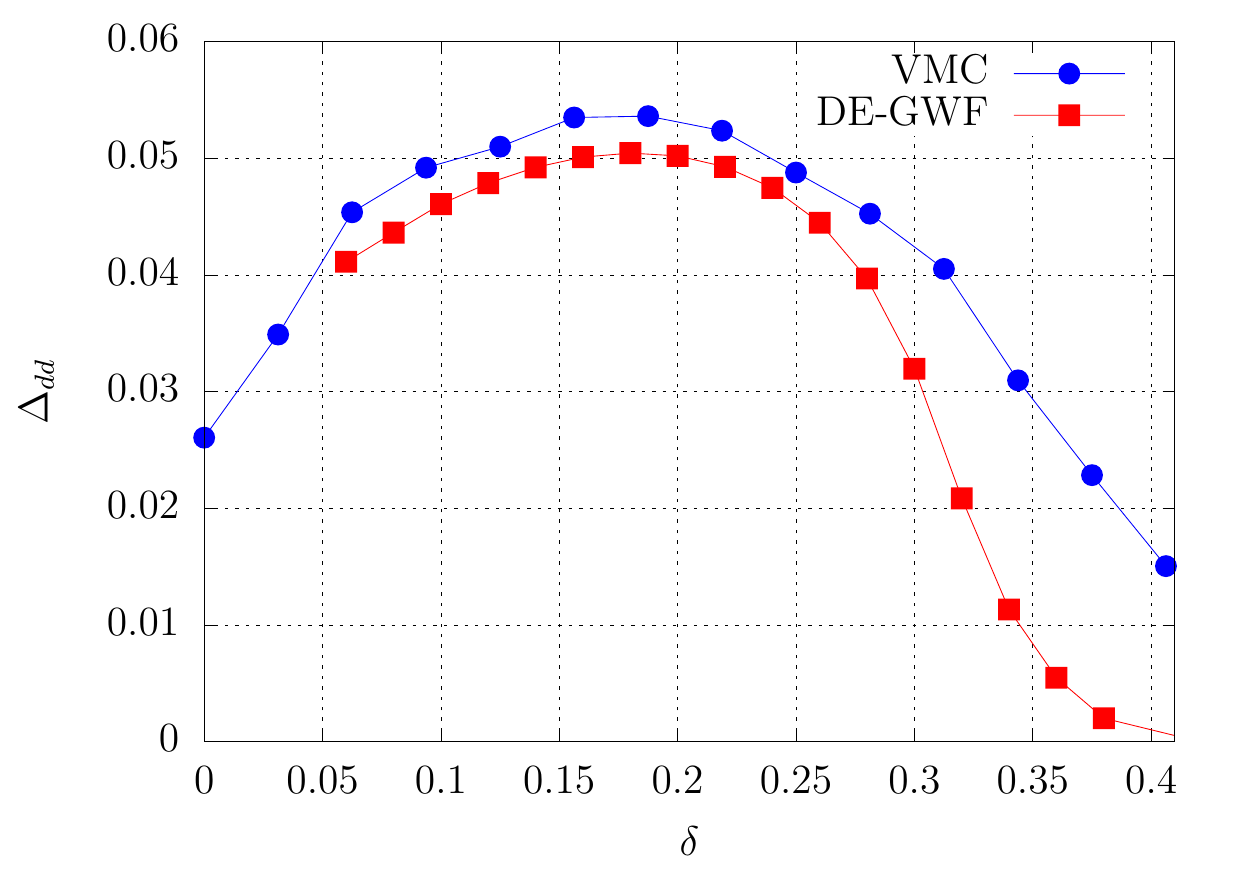}
 \caption{The comparison of $\Delta_{dd}(\delta)$ between DE-GWF~\cite{ZEG1} and VMC approaches. The microscopic parameters are listed in Sec. II.}
 \label{fig:sc_dome_pairs}
\end{figure}

The comparison of $\Delta_{dd}$ for both methods shows that DE-GWF and VMC provide quantitatively similar results (c.f. Fig.~\ref{fig:sc_dome_pairs}), as expected, since both approaches have been supplied with the similar form of variational \emph{ansatz}. It also suggests, that the presence of Jastrow terms in our wave-function  does not affect the solution at least in view of \emph{d-wave} pairing on $d$ orbitals.The discrepancies appearing for higher hole doping are  possibly caused by  the presence of the \emph{finite-size} effects~\cite{GIAMARCHI1} which are absent in the DE-GWF solution (excluding diagrams summations radius in the real space) or optimization issues, as mentioned above. We also observe non-zero pairing amplitudes for $\delta=0$. We estimated previously~\cite{ZEG1}, that $U_d \gtrapprox 13$ eV possibly leads to the full reduction of \emph{d-wave} pairing for the parent compound.                     
The comparison of the results obtained from both methods validates the solution procured for the \emph{assumed form of the wave-function ansatz}. However, characteristics related to the magnetic properties, as well as estimation of charge gap value (presented in the following subsections) may suggest that the adopted form of variational state allows to reproduce the main features of the three-band $d-p$ model.

\subsubsection{Minimal-size real-space \emph{d-wave} pairing operator correlation functions}
Moreo and Dagotto\cite{Moreo}, emphasize that the local $d$-$wave$ operators can provide a more suitable description of the paired holes in cuprates. Their arguments are based on the recent experimental observation of surprisingly small real-space extension \cite{STARFISH} of the Cooper pairs. They also analyse this issue in view of the $p-p$ Cooper pair formation within the single \emph{plaquette}. In this paper we compute the \emph{minimal-size real-space d-wave pairing operator} correlation functions in the framework of VMC method.

\begin{figure}
 \centering
 \includegraphics[width=0.5\textwidth]{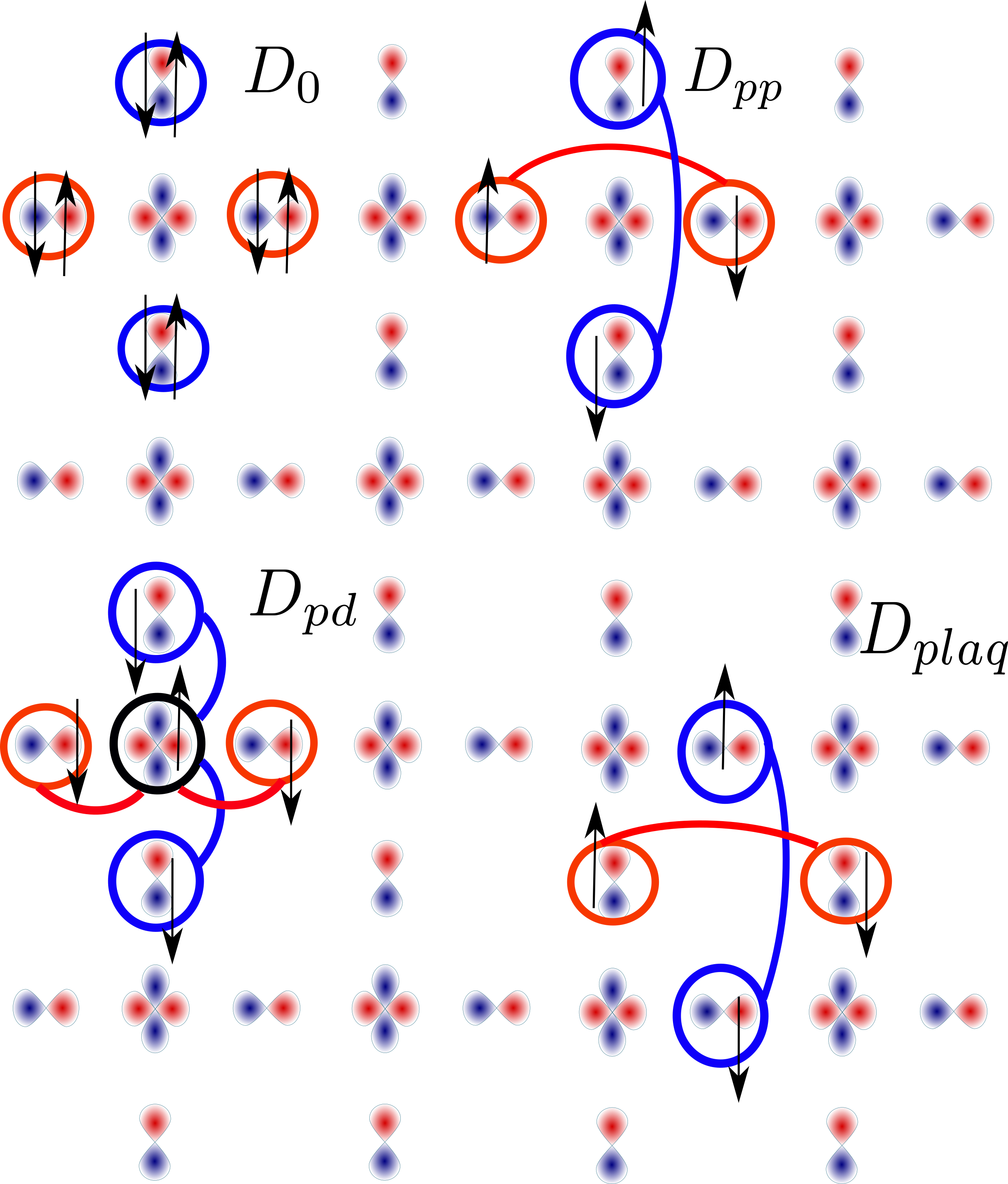}
 \caption{Pairing operators provided in \cite{Moreo} for which the  correlation functions have been computed. Cooper pairs are assigned by connecting lines (excluding $D_0$ for which pair occupies single $p$-orbital). Relative phases signs are marked by colors. }
 \label{fig:scheme_m}
\end{figure}

All of the  four  MSPO preserve $d$-$wave$ symmetry. Intra-site $p$-orbital pair correlation operator is  defined as
\begin{equation}
\label{eq:d0}
   \hat{\Delta}_{D0}^{\dagger}(\mathbf{r})\equiv \sum_{\mu}\gamma_{\mu}\hat{c}_{i(\mathbf{r}+\mathbf{a}_{\mu}/2)\uparrow}^{\dagger}\hat{c}_{i(\mathbf{r}+\mathbf{a}_{\mu}/2)\downarrow}^{\dagger},
\end{equation}
where $\gamma_{\mu}=\text{sgn}(\mu)$.
 The analysis of time evolution of $\hat{\Delta}_{D0}^{\dagger}$ (i.e., Heisenberg equation $-i\frac{d\hat{\Delta}_{D0}^{\dagger}}{dt}=[\hat{H},\hat{\Delta}_{D0}^{\dagger}]$) provides also other pairing operators in the systematic and elegant manner. Namely, the $d$-$p$ pairing operator is given as
 \begin{equation}
 \label{eq:dpd}
    \hat{\Delta}_{Dpd}^{\dagger}(\mathbf{r})\equiv \sum_{\mu,\sigma}f_{\sigma}\gamma_{\mu}\alpha_{i(\mathbf{r}),\mu}\hat{c}_{i(\mathbf{r})\sigma}^{\dagger}\hat{c}_{i(\mathbf{r}+\mathbf{a}_{\mu}/2)\overline{\sigma}}^{\dagger},
\end{equation}
 where $\mu\in\{\pm x,\pm y\}$, and $f(\sigma)=\text{sgn}(\sigma)$ with $\text{sgn}(\sigma)=-\text{sgn}(\overline{\sigma})$, and, $\alpha_{i(\mathbf{r}),\mu}=\pm 1$ consistently with the $d$-$p$ hopping sign convention (cf. Fig.~\ref{fig:Cu_O_hoppings}). Other two MSPO of intra-$p$ type, $\hat{\Delta}_{Dpp}^{\dagger}$ and $\hat{\Delta}_{Dplaq}^{\dagger}$, are also  obtainable in such a procedure and are defined as
  \begin{equation}
  \label{eq:dpp}
    \hat{\Delta}_{Dpp}^{\dagger}(\mathbf{r})\equiv \sum_{\mu,\sigma}f_{\sigma}\gamma_{\mu}\hat{c}_{i(\mathbf{r}+\mathbf{a}_{\mu}/2)\sigma}^{\dagger}\hat{c}_{i(\mathbf{r}-\mathbf{a}_{\mu}/2)\overline{\sigma}}^{\dagger},
\end{equation}
and
   \begin{equation}
   \label{eq:dplaq}
    \hat{\Delta}_{Dplaq}^{\dagger}(\mathbf{r})\equiv \sum_{\mu,\sigma}f_{\sigma}\gamma_{\mu}\hat{c}_{i(\mathbf{r}+\mathbf{a}_{\mu}/2)\sigma}^{\dagger}\hat{c}_{i(\mathbf{r}+\mathbf{a}_{\overline{\mu}}+\mathbf{a}_{\mu}/2)\overline{\sigma}}^{\dagger},
\end{equation}
where $\mu \perp \overline{\mu}$. Schematic representation of the above operators is shown in Fig.\ref{fig:scheme_m}.
It should be noted that these operators are not independent \emph{by construction}. Also, if the ground state  reflects  the \emph{d-wave} superconductivity, simultaneous emergence of all the long-range orderings encoded in Eqs.(\ref{eq:d0})-(\ref{eq:dplaq}) is expected. This important feature can be utilized for the characterization of the system ground state.
We define the CFs of these pairing operators  in the \emph{standard} manner case (cf. Eq.(~\ref{eq:sc_correl})), i.e.,

\begin{equation}
    \label{eq:min_sc_correl}
    D_{D\tau}(\mathbf{R})\equiv \frac{1}{L^2}\sum_{\mathbf{r}}\langle\hat{\Delta}_{D\tau}^\dagger(\mathbf{r}+\mathbf{r})\hat{\Delta}_{D\tau}(\mathbf{r})\rangle
\end{equation}
where $\tau \in \{0,pd,pp,plaq\}$.
 
 \begin{figure}
 \centering
 \includegraphics[width=0.5\textwidth]{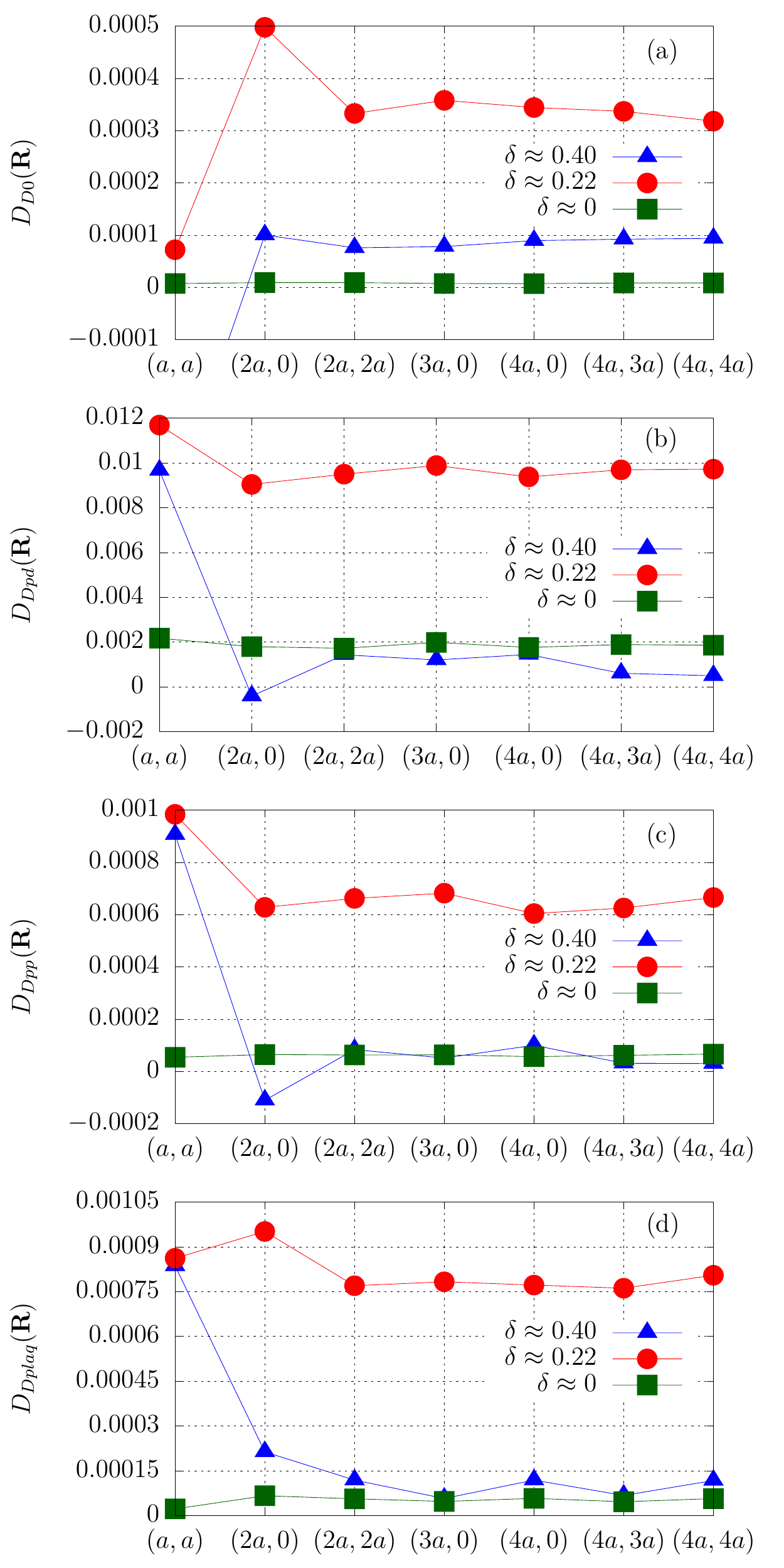}
 \caption{Spatial dependence of the correlation functions $D_{D0}(\mathbf{R})$, $D_{Dpd}(\mathbf{R})$, $D_{Dpp}(\mathbf{R})$, and $D_{Dplaq}(\mathbf{R})$ for  three representative dopings. The amplitude for $\delta\approx0.22$  dominates with increasing $|\mathbf{R}|$ when compared to that for parent compound and high doping cases, as expected.}
 \label{fig:sc_correl_m}
\end{figure}
In Figs. \ref{fig:sc_correl_m}~(a-d) we present spatial dependency of MSPO correlation functions for the representative set of dopings. Disregarding fluctuations originating both from sampling and optimization effects, we observe the saturation of their values within the relatively short distance, i.e., $\mathbf{R}\approx(3a,0)$. Furthermore, all amplitudes fit the picture resulting from \emph{standard} analysis. Namely, the highest amplitude for the most distant pair corresponds to $\delta\approx0.2$, i.e., the optimal doping. Importantly, for the parent compound we obtain a nearly vanishing value of $D_{D0}$ (cf. Fig.~\ref{fig:sc_correl_m}a). The residual non-zero values for $D_{D0}$ and $D_{Dpd}$ at high hole dopings are present, nevertheless they are significantly smaller than for $\delta \approx 0.2$.  

It is reasonable to compare the above amplitudes with the \emph{dominant} $d$-$d$ gap. The order parameter defined in Eq. (\ref{eq:sc_order}) is normalized by $(1/\sqrt{2})^2$ factor, what is not the case for the MSPO correlation functions. Therefore, $D_{Ddd}\equiv2\times D_{\mathbf{R_{max}}}$ should be compared to $D_{D\tau}$ at the $(4a,4a)$ distance. This results in the ratio $D_{Ddd}/D_{Dpd}$ being $\approx 6$ for $\delta\approx 0.2$, which is in a good agreement with results obtained from DE-GWF approach\cite{ZEG1}. Thus $d$-$d$ pairing can indeed be ragarded as the dominant one for the considered form of variational state.

\begin{figure}
 \centering
 \includegraphics[width=0.5\textwidth]{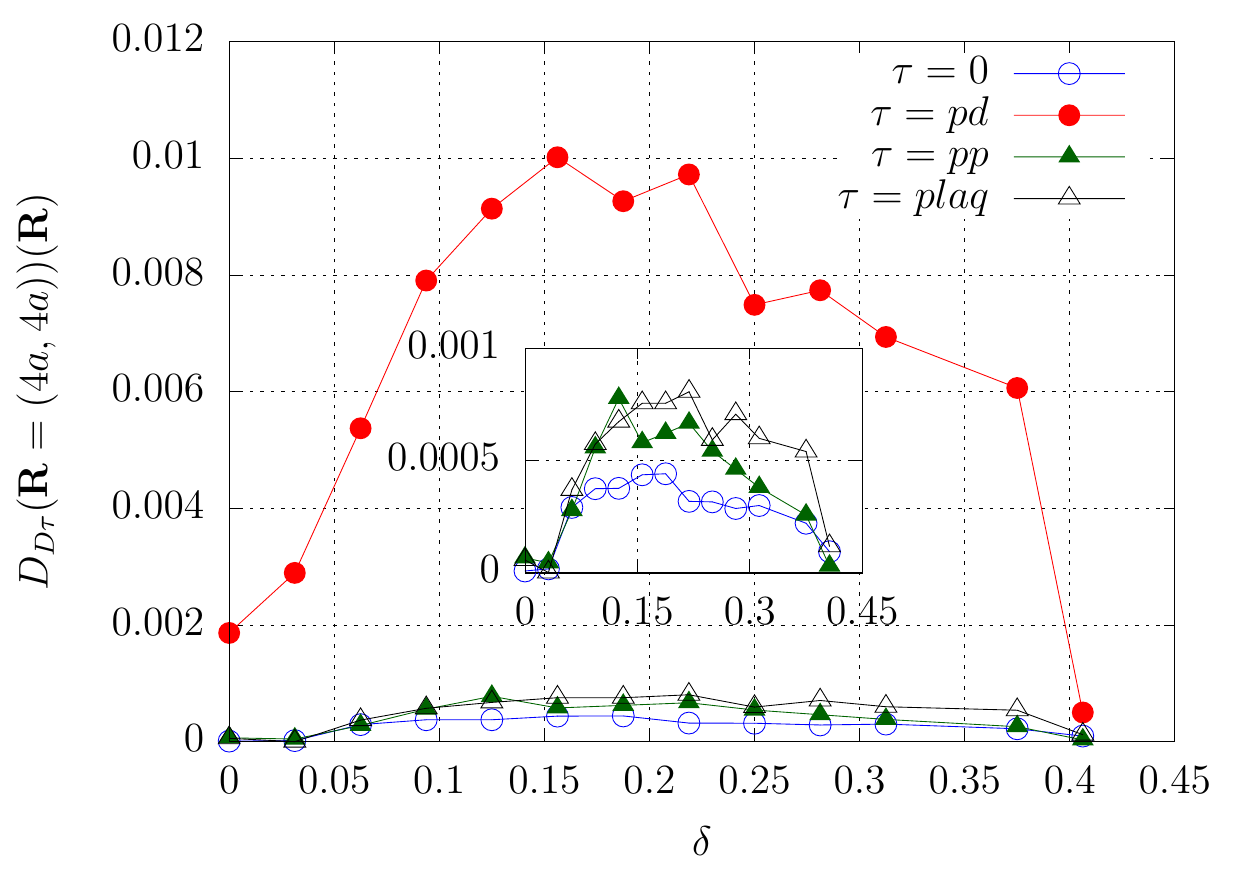}
 \caption{The $D_{D\tau}$ values as function of hole doping obtained for the maximally distanced pairing operators, i.e., $R=(L/2,L2)$. As amplitude of $D_{Dpd}$ dominates by order of magnitude over other CFs, we present dome-like shape for $\tau\in\{0,pp,plaq\}$ in the inset.}
 \label{fig:fig8.pdf}
\end{figure}

In Fig.~\ref{fig:fig8.pdf} we present the values of all MSPO correlation functions for the maximal attainable $R=(4a,4a)$ as a function of hole doping. Note, that we excluded the values for $N_{e}=298 (\delta\approx0.34$) as for all the four CFs considered here the obtained values were unexpectedly high due to optimization issues. As one can see the dome-like shape for all $D_{D\tau}$ is reproduced. Nonetheless, the amplitude of $D_{Dpd}$ is one order of magnitude higher. Detailed analysis of $D_{D0}$, $D_{Dpp}$ and $D_{Dplaq}$  (see inset in Fig.~\ref{fig:fig8.pdf}) provides the evidence of dome-like shape existence in the same range of $\delta$ as observed for $D_{Dpd}$ and $D_{Ddd}$ and thus confirms  the \emph{d-wave} superconducting nature of the ground state. Note that the whole numerical analysis is performed for $S_{tot}^z=0$, so the local diagonal correlations contain Zhang-Rice spin-singlet correlations at local scale.

\subsection{Spin-spin correlations}
The VMC method allows to determine the characteristics of the spin and charge ordering. Though the variational Hamiltonian does not include explicitly antiferromagnetic (AF) terms, the short range correlations of this type can be expected. The existence of the AF order for both hole- and electron-doped cuprates is one of the main features of their phase diagram. Therefore, we investigate if the considered \emph{ansatz} is able to reproduce such a tendency.

\begin{figure}
 \centering
 \includegraphics[width=0.5\textwidth]{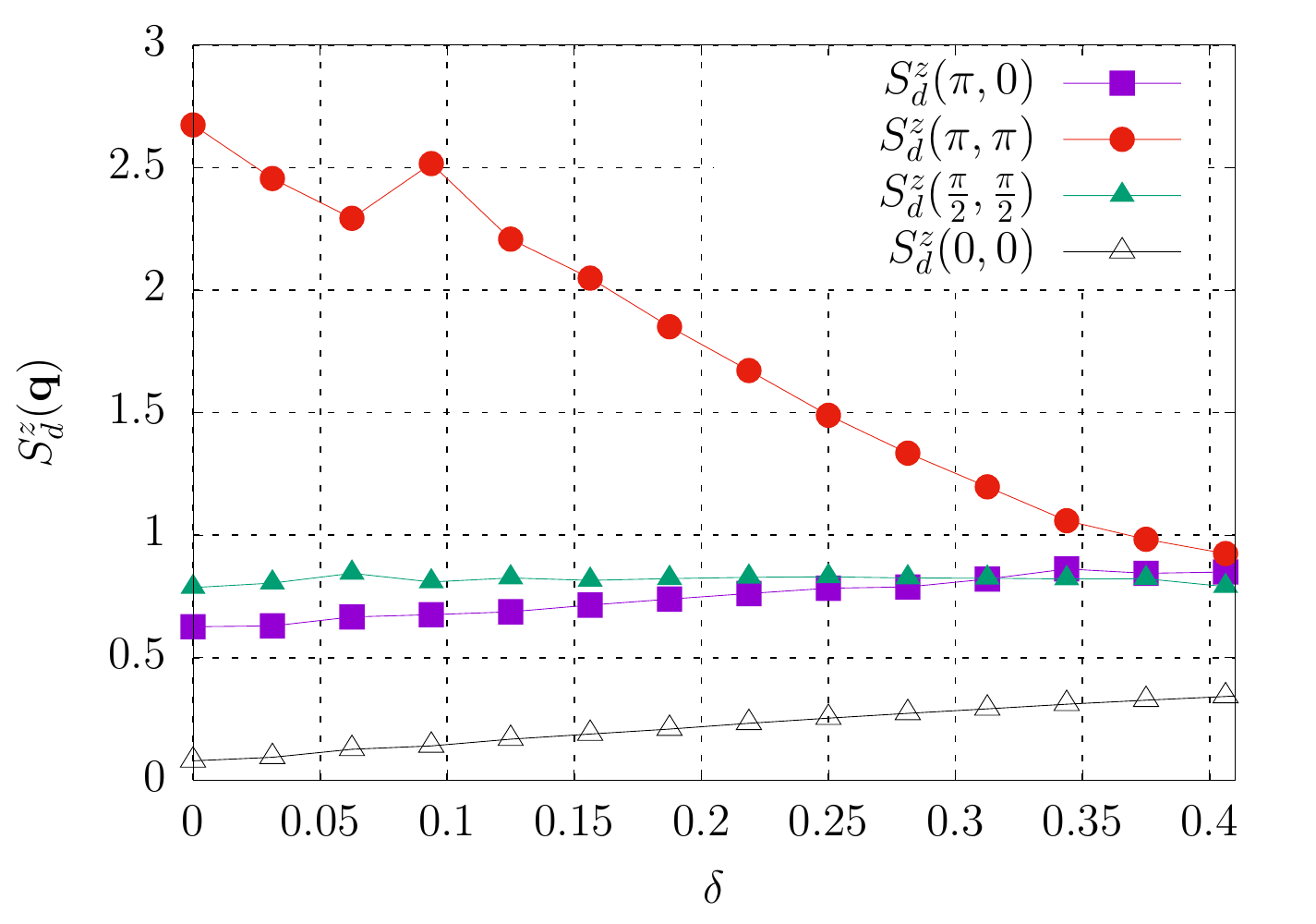}
 \caption{Static spin-spin susceptibilities as a function of hole-doping $\delta$ for the $d$ orbitals. $S_d^{z}(\pi,\pi)$ is dominant, particularly with decreasing $\delta$, and attains maximum value for the parent compound.}
 \label{fig:ss_d_correls}
\end{figure}

We perform the analysis of $z$-component of spin-spin correlation functions defined in a standard manner\cite{QMC_3band_2016, TOCCHIO1, ZHAO1}, namely,
\begin{equation}
    S_{l}^{z}(\mathbf{R})=\frac{1}{L^2}\sum_{i,\mathbf{R}}\left\langle(\hat{n}_{il\uparrow}-\hat{n}_{il\downarrow})(\hat{n}_{j(\mathbf{R})l\uparrow}-\hat{n}_{j(\mathbf{R})l)\downarrow})\right\rangle
\end{equation}
and the static spin-spin susceptibility, which has the form
\begin{equation}
    S_{l}^{z}(\mathbf{q})=\sum_{\mathbf{R}}e^{i\mathbf{q}\cdot\mathbf{R}}S_{l}(\mathbf{R}),
\end{equation}
with $\mathbf{q}$ being the ordering vector given in $1/a$ units.
In Fig.~\ref{fig:ss_d_correls} we present $S_l^z(\mathbf{q})$ for vectors $\mathbf{q}=\{(\pi,0),(\pi,\pi),(\frac{\pi}{2}),\frac{\pi}{2}),(0,0)\}$.
As one can see, the amplitude for the ordering vector $\mathbf{q}=(\pi,\pi)$ dominates over others as the hole doping is reduced in the system. This corresponds to the tendency of establishing the magnetic state with \emph{staggered} magnetization, at least at short range. Note, that our result is in the quantitative agreement with that obtained  by means of the determinant quantum Monte Carlo DQMC method~\cite{QMC_3band_2016}, e.g., value of $S_d^z(\pi,\pi)$ at $\delta=0$ is  $\approx2.6$. However, our result refers to that procured for a larger system. Thus the ratio $\frac {S_d^z(\pi,\pi)}{L^2}$ fits the \emph{finite-size} scaling analysis performed by Kung et al.\cite{QMC_3band_2016}. In our solution the absence of the long-range AF order is expected due to no AF terms in $\hat{\mathcal{H}}_{eff}$, whereas in DQMC, where $T>0$, it originates from the Mermin-Wagner theorem. Despite these circumstances both approaches reproduce similar spin physics.
\begin{figure}
 \centering
 \includegraphics[width=0.5\textwidth]{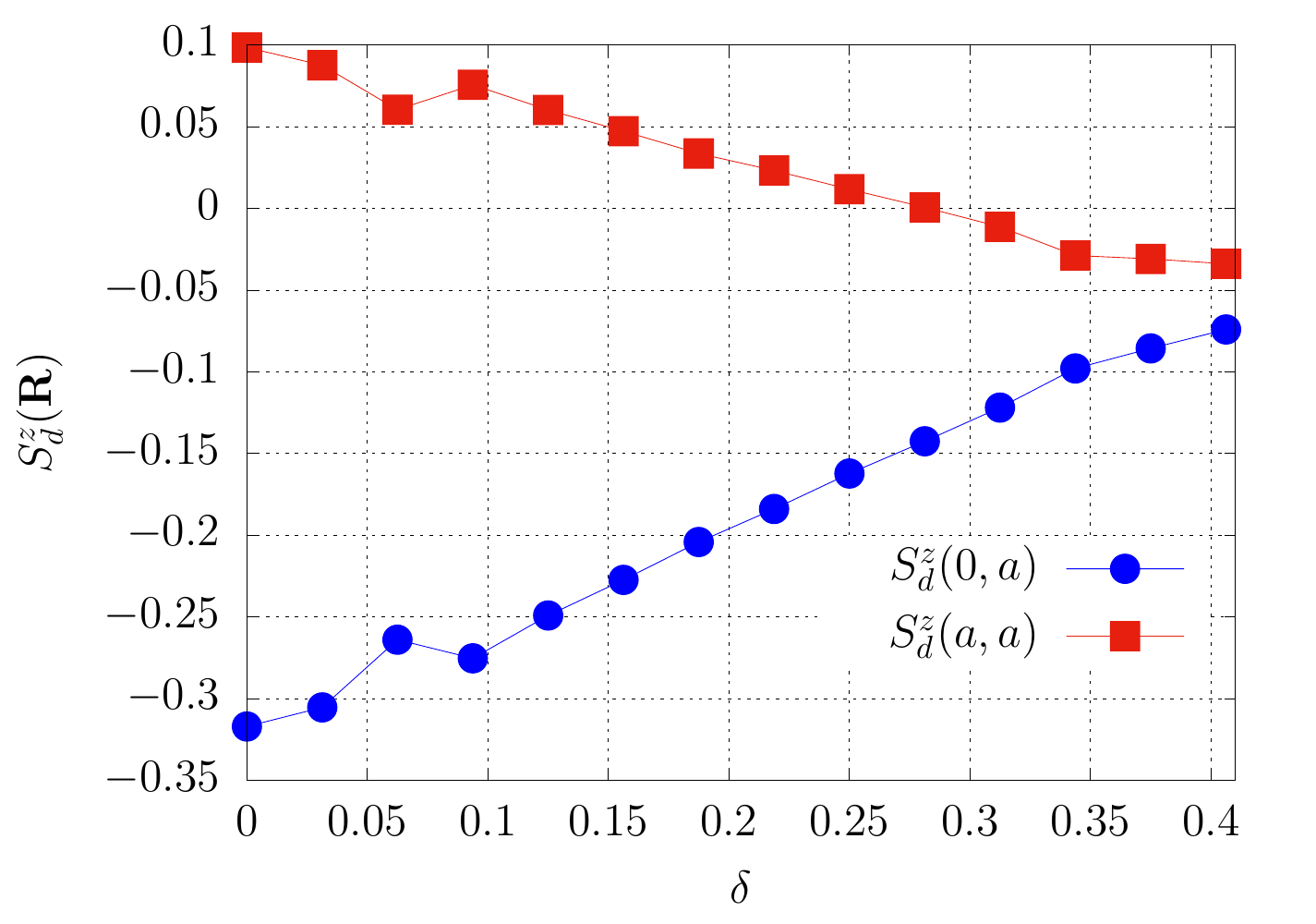}
 \caption{The doping dependence of the correlation functions $S_d^z(\mathbf{R})$ for the nearest- and next-nearest $d$ orbitals. The maximal absolute values correspond to $\delta=0$, and their signs correspond to the development of staggered magnetization.}
 \label{fig:corecomp}
\end{figure}
\begin{figure}
 \centering
 \includegraphics[width=0.5\textwidth]{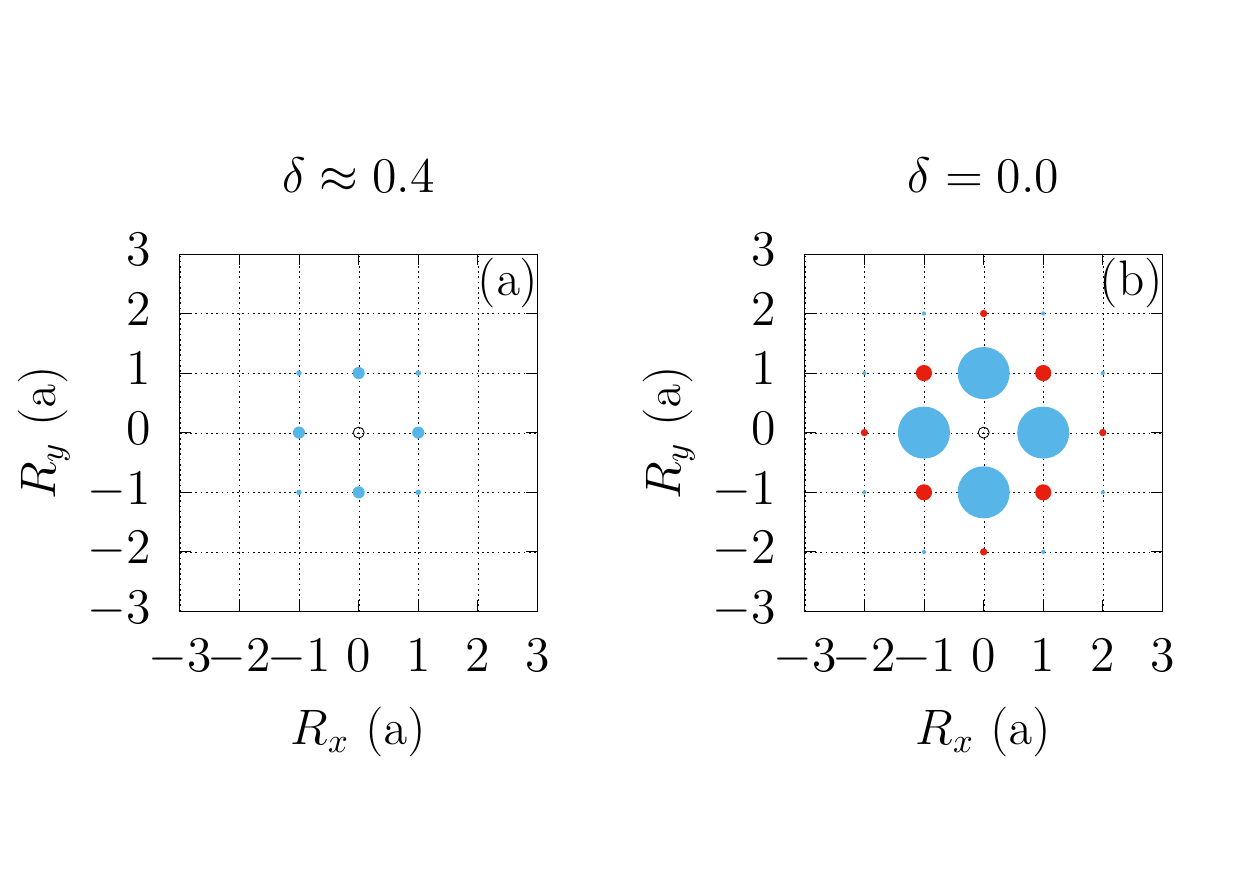}
 \caption{The decay of spin-spin spatial correlations in real space for $\delta\approx0.4$ (a) and $\delta=0$ (b). The radius of the circles is proportional to the value of $S_d^z(\mathbf R)$; the color indicates the sign of the amplitude: positive (red) and  negative (blue). For the sake of brevity we exclude the auto-correlation function (central black dot). }
 \label{fig:farss}
\end{figure}
The evidence of the AF correlations enhancement with decreasing hole-doping manifests itself also in terms of the real-space  analysis. In Fig.\ref{fig:corecomp} we present spatial correlation function $S_d^z(0,a)$ and $S_d^z(a,a)$. As one can see, with decreasing hole-doping, nn. orbitals are occupied by the antiparallel spins, whereas correlations between next-nn (nnn) becomes positive, indicating parallel  orientation of $z$-component of the further spins. In spite of the fact that the spin-spin correlations are short-ranged, in the  vicinity of $\delta=0$ they decay slowly with the distance   (c.f. Fig.~\ref{fig:farss}). In the analyzed doping  range we have not found any indication of cross-over from AF to ferromagnetic correlations. Namely, $S_d^z(0,a)$ remains negative and increases with increasing hole-doping.

\begin{figure}
 \centering
 \includegraphics[width=0.5\textwidth]{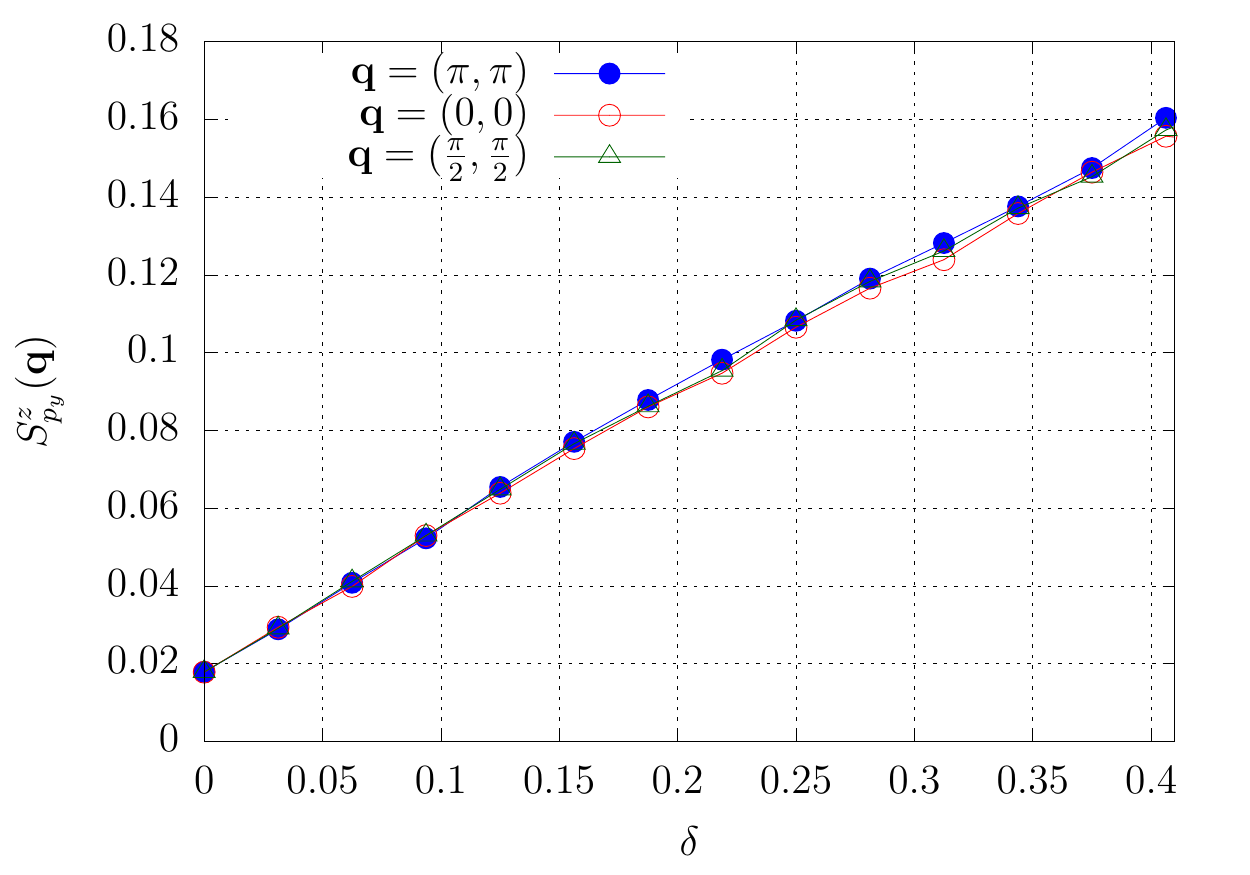}
 \caption{Equal-time correlation functions $S_{p_y}^z(\mathbf{q})$ for the selected wave vectors as a function of doping. In the considered range of doping the system does not exhibit spin ordering at the $p$-orbitals.
 }
 \label{fig:ss_py}
\end{figure}

For the sake of completeness, the correlation functions related to $p$ orbitals are presented in Fig.~\ref{fig:ss_py}. As one can see, there is no particular spin-order for each selected wave-vectors. Values of $S_{p_{y}}^z({\mathbf{q}})$ decrease monotonically with decreasing hole-doping. This result agrees with that obtained by means of DQMC\cite{QMC_3band_2016}.

\subsection{Charge gap}
As cuprates fall into the class of strongly correlated systems, the emergence of electron-electron induced insulating phase is characteristic to these compounds. The outcome  of our estimation is directly comparable to other theoretical treatments as well as to the experimental results.

One of the methods for calculating the charge gap (CG) $\Delta_{CG}$ (which identifies insulating state) is based on the Single-Mode Approximation (SMA), which has been proved to be an efficient method for the Hubbard-type systems. Within such analysis one has to determine quantity
\begin{equation}
\Delta_{CG} \propto \lim_{\mathbf q \rightarrow 0}\frac{\chi^{c}(\mathbf{q})}{\mathbf{q}},    
\end{equation}
where $\chi^{c}(\mathbf{q})$ is the Fourier transform of equal-time charge-charge correlation function. Unfortunately, the minimal norm of the wave vector for $L=8$ is  $|\mathbf{q}|=\frac{\pi}{2}$, thus we are not able to provide a firm estimate of $\Delta_{CG}$ along these lines. Instead, we determine the value of $\Delta_{CG}$ in a standard manner. Namely,
\begin{equation}
\Delta_{CG}\approx\frac{2E(N_{e})-E(N_{e}+2)-E(N_{e}-2)}{2},
\end{equation}
where $E(N_{e})$ is the total energy of the system at the doping value corresponding to particular number of electrons $N_{e}$. The above formula previously used by us in a different context~\cite{Biborski}, can be applied  directly here due to the fact the in this analysis one can safely assume that $S^{z}_{tot}=0$, and $\Delta N_{e}=2$.

\begin{figure}
 \centering
 \includegraphics[width=0.45\textwidth]{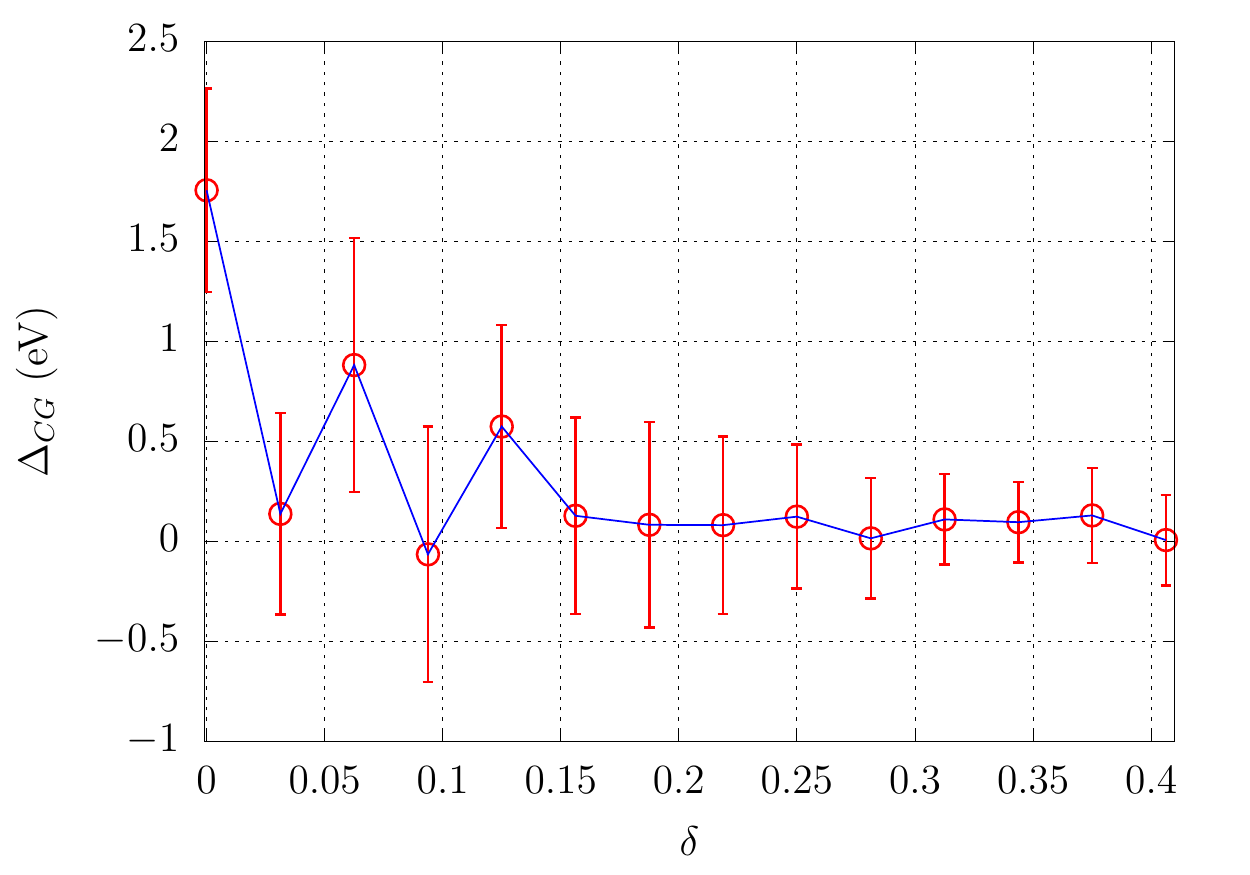}
 \caption{Charge gap  $\Delta_{CG}$ versus $\delta$. The  gap emerges for $\delta\lessapprox0.1$ attaining, its maximum for the parent compound ($\delta=0$).
 }
 \label{fig:chargegap}
\end{figure}

In Fig.\ref{fig:chargegap} we present $\Delta_{CG}$ as a function of $\delta$. For $\delta \gtrapprox 0.12$ we still obtain a small but non-zero values of $\Delta_{CG}$, which should be considered as residual and not identified as an indicator of the insulating state. Close to the zero doping, we obtain the maximal value of $\Delta_{CG}(\delta=0)\approx 1.78$ eV, as expected for the parent compound which agrees well with  those reported in experiments~\cite{Falck,Uchida,Tokura,CooperS,Terashigeeaav2187}, i.e., $\Delta_{CG}\approx 1.32-2.2$ eV for the group of layered structure compounds $\text{X-CuO}_{2}$, where X refers to  lanthanide (La, Sr, Nd, Ca, Sm, Tb).
Kung et al.~\cite{QMC_3band_2016} reports the value of indirect gap $\approx 0.77$ eV (after the extrapolation to zero temperature). The authors discuss if such a low value - when compared to the experiment  - originates from \emph{finite-size} effects or is connected with temperature extrapolation issues. The maximal size cluster taken for that study was $6 \times 6$, i.e. (smaller than the one examined by us) as well as $2 \times 2$ clusters were treated at $T=0$ also in the framework of cluster perturbation theory (CPT) and ED\cite{QMC_3band_2016}. The latter method provided $\Delta_{CG}\approx 1.7$, which is very close to the value obtained by us. This may suggest that the extrapolation to $T=0$ for data obtained in the framework of DQMC \cite{QMC_3band_2016} affected value of $\Delta_{CG}$, thus finite system size effects seem  not to be decisive in this matter. This issue  needs a further analysis, since we do not have a systematic analysis  of \emph{finite-size} effects.

\section{Conclusions and outlook}
In this paper we have considered a three-band $d-p$ model of the copper-oxygen plane within the  VMC approach with a wave-function \emph{ansatz} containing both onsite Gutzwiller and intersite Jastrow correlators  in  real space. The analysis of superconducting pairing properties in view of the so-called \emph{standard} analysis, i.e., the one based on $d$-orbital pairing correlation functions, provided us with results which are  quantitatively consistent with our previous work, as well as, qualitatively with selected experimental observations. As an extension of our previous work we have also calculated the spatial distribution functions of pairing operators proposed by Moreo and Dagotto in their very recent report \cite{Moreo}. We found that the amplitude of the correlation function (CF) is the highest (order of magnitude higher) for the $D_{pd}$ operator when compared to those consisting of $p$ orbitals only. Moreover, the  considered CFs show the \emph{dome-like} behavior as a function of hole-doping, which is similar to the nearest-neighbor $d$-$d$ pairing amplitude. According to our study the correlation functions for SMPO parameters can be regarded as convenient observables for the characterization of the \emph{d-wave} paired state in the $d$-$p$ model. Recapitulating, scrutinization of pairing observables in the context of this paper, as well as, analysis performed recently\cite{ZEG1}, indicate that both inter- and intra- orbital pairing amplitudes are responsible for the net \emph{d-wave} superconductivity in the three-band $d$-$p$ model. Nonetheless, the dominant contribution to the superconducting state results from the $d$-$d$ pairing\cite{ZEG1}. 

For the sake of completeness, we have also determined \emph{spin-spin} equal-time correlation functions. Even though the utilized \emph{ansatz} is not supplemented with explicit antiferromagnetic terms, we have observed short range AF ordering on the $d$-orbitals. Detailed analysis, brought us to conclusions similar to those obtained by Kung et al. \cite{3band_VMC_Mott}. Particularly, static \emph{spin-spin} susceptibilities agree quantitatively with the DMC solution. Moreover, the estimated value of charge-gap is  $\Delta_{CG}\approx 1.78 \text{ eV}$, which fits surprisingly well experimental data \cite{Falck,Uchida,Tokura,CooperS,Terashigeeaav2187}. 

Recapitulating, we have retrieved the main features of hole-doped cuprate compounds by means of VMC method, within  compact, Gutzwiller-Jastrow variational approach. According to experimental findings\cite{Comin}, the symmetry of charge order in the cuprates is likely to be complex, and the role of $p$-orbitals is supposed  to be quite important. Nevertheless, we have not analyzed the onset of \emph{charge} ordering\cite{Comin}. This issue can be related both to the supercell size, as well as to the form of variational \emph{ansatz}. Possibly, the application of most general, Pfaffian-wave-function and more distant Jastrow terms, with a minimal dose of symmetries, could provide a better understanding of this state. However, in such a scenario the number of variational parameters is large and the optimization procedure may become too complex.  The recent development of  dedicated VMC codes may help to overcome these difficulties\cite{MISAWA2019447}, potentially even at the \emph{ab-initio} level \cite{nakano2020turborvb,PhysRevB.101.045124}. We should be able to see a progress in this matter in the near future.

\section{Acknowledgement}

 A.B. and M.Z. acknowledge the financial support through the Grant SONATA, No. 2016/21/D/ST3/00979 from the National Science Centre (NCN), Poland. J.S.  acknowledges  the  financial  support  by the Grant OPUS No. UMO-2018/29/B/ST3/02646 from the National Science Centre (NCN), Poland. This work is supported in part by  the computing PL-Grid Infrastructure. A.B. would also like to thank Stefan Siekanka for his help in the improvement of the computational code.

\appendix

\bibliography{3band.bib}

\end{document}